\documentclass{aa_v7.0}
\usepackage[varg]{txfonts}
\usepackage{graphicx}

\newcommand{\kmprs}  {\mbox{\rm km\,s$^{-1}$}}

\newcommand{\feh} {\mbox{\rm [Fe/H]}}

\newcommand{\ch} {\mbox{\rm [C/H]}}
\newcommand{\oh} {\mbox{\rm [O/H]}}
\newcommand{\co} {\mbox{\rm [C/O]}}
\newcommand{\ofe} {\mbox{\rm [O/Fe]}}

\newcommand{\nife} {\mbox{\rm [Ni/Fe]}}

\newcommand{\teff}  {\mbox{$T_{\rm eff}$}}
\newcommand{\logteff} {\mbox{${\rm log}\,T_{\rm eff}$}}
\newcommand{\logg}  {\mbox{{\rm log}\,$g$}}
\newcommand{\turb}  {\mbox{$\xi_{\rm turb}$}}

\newcommand{\HII} {\ion{H}{ii}}

\newcommand{\CI} {\ion{C}{i}}
\newcommand{\OI} {\ion{O}{i}}

\newcommand{\FeI} {\ion{Fe}{i}}
\newcommand{\FeII} {\ion{Fe}{ii}}
\newcommand{\NiI} {\ion{Ni}{i}}

\newcommand{\Mv} {\mbox{$M_V$}}

\newcommand{\VK}{\mbox{($V\!-\!K)$}}

\newcommand{\by}{\mbox{($b\!-\!y)$}}

\def\ltsima{$\; \buildrel < \over \sim \;$}
\def\simlt{\lower.5ex\hbox{\ltsima}}
\def\gtsima{$\; \buildrel > \over \sim \;$}
\def\simgt{\lower.5ex\hbox{\gtsima}}
\begin{document}

\title{The carbon-to-oxygen ratio in stars with planets
\thanks{Based on data products from observations made with ESO Telescopes
at the La Silla Paranal Observatory under programmes given in Table \ref{table:obs}.}
\fnmsep\thanks{Tables \ref{table:obs} and \ref{table:results} are provided as online material
and is available in electronic form at {\tt http://www.aanda.org}.}}

\author{P. E. Nissen \inst{}}

\institute{Stellar Astrophysics Centre, 
Department of Physics and Astronomy, Aarhus University, Ny Munkegade 120, DK--8000
Aarhus C, Denmark.
\email{pen@phys.au.dk}}

\date{Received 5 February 2013 / Accepted 2 March 2013}

\abstract
{In some recent works, the C/O abundance ratio in high-metallicity stars with planets
is found to vary by more than a factor of two, i.e. from 
$\sim \! 0.4$ to C/O$\simgt 1$. This has 
led to discussions  about the existence of terrestrial planets with a carbon-dominated 
composition that is very different from the composition of the Earth.}
{The mentioned C/O values were obtained by determining carbon abundances
from high-excitation \CI\ lines and oxygen abundances from the forbidden [\OI ] line
at 6300\,\AA . This weak line is, however, strongly affected by a nickel blend 
at high metallicities. Aiming for more precise C/O ratios,
oxygen abundances in this paper are derived from the 
high-excitation \OI\ triplet at 7774\,\AA .}
{The \CI\ lines at 5052 and 5380\,\AA\ in HARPS spectra were applied to
determine carbon abundances of 33 solar-type stars for which  
FEROS spectra are available for determining 
oxygen abundances from the \OI\ $\lambda 7774$ triplet.  
Differential abundances with respect to the Sun were derived 
from equivalent widths using MARCS model atmospheres. Non-LTE
corrections were included, and the analysis was carried out with
both spectroscopic and photometric estimates of stellar effective temperatures
and surface gravities.} 
{The results do not confirm the high C/O ratios previously found.
C/O shows a tight, slightly increasing dependence on metallicity,  
i.e. from C/O $\simeq 0.58$ at $\feh = 0.0$ to
C/O $\simeq 0.70$ at $\feh = 0.4$ with an rms scatter of only 0.06.}  
{Recent findings of C/O ratios higher than 0.8 in
high-metallicity stars seem to be spurious due to statistical errors
in estimating the strength of the weak [\OI ] line in the  
\NiI\ blended $\lambda 6300$ feature. 
Assuming that the composition of a proto-planetary disk is the same
as that of the host star,
the C/O values found in this paper lend no support to the
existence of carbon-rich planets. The small scatter of C/O
among thin-disk stars
suggests that the nucleosynthesis products
of Type II supernovae and low- to intermediate-mass stars  
are well mixed in the interstellar medium.
}


\maketitle

\newpage

\section{Introduction}
\label{sect:introduction} 
Recent high-resolution spectroscopic surveys suggest 
that the carbon-to-oxygen abundance ratio, C/O\footnote{C/O 
is defined as $N_{\rm C}/N_{\rm O}$,
where $N_{\rm C}$ and $N_{\rm O}$ are the number densities of
carbon and oxygen nuclei, respectively. It should not
be confused with the solar-normalized logarithmic ratio, [C/O] $\equiv 
{\rm log}(N_{\rm C}/N_{\rm O})_{\rm star}\,\, - \,\,{\rm log}(N_{\rm C}/N_{\rm O})_{\rm Sun}$.},
in the atmospheres of metal-rich F and G dwarf stars
varies by more than a factor of two. Delgado Mena et al.
(\cite {delgado10}) find a range in C/O from
$\sim \! 0.4$ to $\simgt 1$ for 370 stars observed with
HARPS or CORALIE, including about 100 stars with detected planets.
A similar range in C/O has been found by Petigura \& Marcy
(\cite{petigura11}) for about 700 stars observed with HIRES
of which 80 have detected planets. Even if one corrects
the distribution of C/O for the quoted observational errors,
there seems to
be a small fraction (1-5\%) of the stars having C/O $\simgt 1$, and a significant
fraction (10 - 15\,\%) having C/O\,$> 0.8$ (Fortney \cite{fortney12})
in contrast to the solar ratio of C/O\,$\simeq 0.55$.

Assuming that the composition of a proto-planetary disk is the same
as that of the host star, variations in the C/O abundance ratio 
among solar-type stars may have interesting effects on the composition and
structure of planets. Thus, Kuchner \& Seager (\cite{kuchner05}) discuss
the possible existence  of extrasolar ``carbon planets" consisting of
carbides and graphite instead of ``Earth-like" silicates.
When modelling terrestrial planet formation, Bond et al. (\cite{bond10})
find that C/O\,$> 0.8$ leads to the formation of such carbon planets. 
A particularly interesting case is super-earth 55 Cancri e.
Studies of its transit across the host star (Winn et al. \cite{winn11};
Demory et al. \cite{demory11}) lead to a radius $R \simeq 2.0 \, R_{\oplus}$
and a mass $M \simeq 8.4 \, M_{\oplus}$. Adopting an oxygen-rich composition,
a massive water envelope over an Earth-like interior is required to
explain the observed radius. Madhusudhan et al. ({\cite{madhusudhan12})
have, however, shown that the mass and radius can also be explained by 
a carbon-rich interior without a volatile envelope. Interestingly, the
star hosting this planet, i.e. 55 Cancri alias HD75732, has
C/O = 1.12 according to Delgado Mena et al. (\cite {delgado10}).
    
The existence of dwarf stars with C/O\,$> 1$ has
been questioned by Fortney (\cite{fortney12}). Among cool K and M dwarfs,
C/O\,$> 1$ leads to strong C$_2$ and CN bands that are easy to detect. 
From various large spectroscopic surveys, including the SDSS,
Fortney notes that the relative frequency of such ``carbon" stars
is less than $\sim \! 10^{-3}$, and discusses various reasons that the C/O
ratio may have been overestimated in solar-type stars.

In the works of Delgado Mena et al.  (\cite {delgado10}) and
Petigura \& Marcy (\cite{petigura11}), carbon abundances were determined
from high-excitation \CI\ lines
and oxygen abundances from the zero-excitation, 
forbidden [\OI ] line at 6300\,\AA . The large difference in 
excitation potential makes the derived C/O sensitive to errors
in stellar effective temperature and to the temperature
structure of the model atmospheres. Although the analysis in the
two works was made differentially to the Sun, the use of homogeneous
1D model atmospheres might have introduced systematic errors of C/O as a function
of metallicity. Furthermore, the [\OI ] line is overlapped by a \NiI\ line, 
which is calculated 
to have a strength of about 55\% of the [\OI ] line in the solar flux spectrum
(Caffau et al. \cite{caffau08}), if one adopts the $gf$-value 
determined by Johansson et al. (\cite{johansson03}).
Given that [O/Fe] decreases and [Ni/Fe] increases with increasing [Fe/H],
(e.g. Bensby et al. \cite{bensby05}), the \NiI\ blend becomes more important 
at the high metallicities for which most of the high C/O values have been found. 
This makes it difficult to determine precise oxygen abundances from
the [\OI ] line at high metallicities, and systematic errors in [O/H]
are introduced if the \NiI\ blend is not correctly calibrated for the 
solar spectrum. Finally, the measured equivalent width of the weak [\OI ]
line depends critically on the continuum setting, and sometimes it is
difficult to correct for overlapping telluric O$_2$ lines.

Instead of the [\OI ] line, the \OI\ triplet at 7774\,\AA\ may
be used to determine oxygen abundances. These three lines have a high
excitation potential like the \CI\ lines, which makes
the derived C/O ratio insensitive to errors in effective temperature
and the temperature structure of the model atmospheres. Furthermore, the
lines are practically unblended. A drawback is that the triplet
is affected by deviations from local thermodynamic equilibrium (non-LTE),
(Kiselman \cite{kiselman01}), which is not the case for the 
forbidden [\OI ] line. 

The ESO 3.6m HARPS spectra (Mayor et al. \cite{mayor03})
used by Delgado Mena et al.  (\cite {delgado10})
have a maximum wavelength of 6900\,\AA , but for a good fraction 
of their stars, ESO 2.2m FEROS spectra (Kaufer et al. \cite{kaufer99}) covering the 
\OI\ triplet are available. In this paper, these FEROS spectra
have been used to derive oxygen abundances from the triplet together
with carbon abundances from the \CI\ lines in the HARPS spectra.
Sect. 2 contains a description of the stellar spectra and the measurement 
of equivalent widths. The model-atmosphere 
analysis, including non-LTE corrections, is dealt with in Sect. 3, and
the abundances obtained when adopting
spectroscopic and photometric estimates of 
\teff\ and \logg , respectively, are presented in Sect. 4 and 5.
Finally, Sect. 6 provides a comparison with 
other works and a discussion of the Galactic dispersion and evolution 
of C/O.

\section{Stellar spectra and equivalent widths}
\label{sect:EWs}

Tables 1 and 2 in Delgado Mena et al. (\cite{delgado10}) contain
a list of 100 dwarf stars with detected planets for which they have determined
C and O abundances based on spectra from the HARPS GTO survey.
Out of this sample, 38 stars
have FEROS spectra available in the ESO Science Archive. 
Of these, five stars have \teff\ around 5100\,K.
The remaining 33 stars (listed in Table \ref{table:obs})
have $5400 < \teff < 6400$\,K, which is a suitable range for a differential abundance
analysis with respect to the Sun. 

Pipeline extracted and wavelength calibrated spectra observed under 
the programmes listed in Table \ref{table:obs} were acquired from the 
ESO Science Archive. The HARPS spectra cover a wavelength range
from 3800 to 6900\,\AA\ and have a resolution of 
$R\simeq \! 115\,000$. After combination of many individual spectra,
the signal-to-noise ($S/N$) ratio ranges from 250 to more than 1000.
The FEROS spectra range from 3500 to 9200\,\AA\  
with a resolution of $R\simeq \! 48\,000$ . 
Most of these spectra have $S/N$ from 200 to 300 (in the region of
the \OI\ triplet), except for \object{HD\,216770} that has $S/N = 120$. 

The spectra were normalized with the IRAF {\tt continuum} task using a cubic spline
fitting function  of relatively low order, i.e. a wavelength scale $> \! 20$\,\AA .
The measurement of an equivalent width ($EW$) was carried out with reference
to local continuum regions selected to be free of lines in the solar spectrum
and situated within 4\,\AA\ from the line. Care was taken to use the same
continuum windows in all stars.

The measurement of the equivalent widths of the \OI\ triplet lines 
causes particular problems.
These lines are somewhat strong and have very broad wings. If measured by
fitting a Voigt profile, the equivalent width becomes very sensitive
to the setting of the continuum. This is even a problem in solar spectra
having extremely high resolution and $S/N$. Thus, the $EW$s of the \OI\ triplet lines
measured by Asplund et al. (\cite{asplund04}) and Caffau et al. ({\cite{caffau08})
differ by 5 to 10\,m\AA , which is a major reason for the
difference in the solar oxygen abundance derived in these two 3D studies.
In the present work, equivalent widths are, instead, measured by Gaussian fitting  
using the IRAF {\tt splot} task, which for the resolution of FEROS
turns out to provide nearly the same $EW$ as 
obtained by direct integration over a range of $\pm 0.4$\,\AA\ centred on the
line. The method  neglects the contribution from the broad wings, but
this was compensated for by also calculating the $EW$ in the model-atmosphere analysis
over a window of $\pm 0.4$\,\AA . The same procedure was
followed for the two \CI\ lines, although
they are sufficiently weak to ensure that Gaussian
fitting provides a good estimate of the total $EW$ of the line. 

The measured equivalent widths are given in Table \ref{table:obs}. 
It is noted that weak CN and C$_2$ lines are present in the region
of the \OI\ triplet, but fitting with a Gaussian profile
minimizes the influence of such blends. In general, the equivalent widths
of the \OI\ lines have a precision of 2 -- 3\,m\AA\ and those  
of the \CI\ lines a precision of about 1\,m\AA .

To make a reliable differential analysis with respect to the Sun,
it is important to apply a solar flux spectrum observed in the same
way as the stellar spectra. Fortunately, this is possible; spectra of
reflected sunlight from Ceres and Ganymede have been obtained with both
HARPS and FEROS, and with $S/N$ ratios similar to those of the programme stars.
As seen from  Table \ref{table:obs}, the equivalent widths measured
for the two minor planets agree very well. Therefore, the average values have 
been applied in the model-atmosphere analysis.

\onltab{1}{
\begin{table*}
\caption[ ]{ESO observing programme numbers, $S/N$ ratios,
and equivalent widths measured by Gaussian fitting to line profiles.}
\label{table:obs}
\setlength{\tabcolsep}{0.20cm}
\begin{tabular}{lccccrrrrr}
\noalign{\smallskip}
\hline\hline
\noalign{\smallskip}
     &  \multicolumn{2}{c}{HARPS}  &  \multicolumn{2}{c}{FEROS} & \CI \, $\lambda 5052$ &
\CI \, $\lambda 5380$ & \OI \, $\lambda 7772$ & \OI \, $\lambda 7774$ & \OI \, $\lambda 7775$    \\
  ID & Programme  &  $S/N$  &  Programme & $S/N$ & $EW$(m\AA )&
$EW$(m\AA )& $EW$(m\AA ) & $EW$(m\AA ) & $EW$(m\AA ) \\
\hline
\noalign{\smallskip}
HD\,142     &  72.C-0488 &  450 & 83.A-9011 & 300 &  69.8 &  45.2 & 131.7 & 114.3 &  95.8  \\
HD\,1237    &      -     &  350 & 60.A-9700 & 200 &  30.9 &  16.8 &  59.1 &  50.2 &  36.4  \\
HD\,4308    &      -     &  600 & 74.D-0086 & 500 &  25.5 &  14.1 &  68.1 &  58.5 &  42.4  \\
HD\,16141   &      -     &  300 & 83.A-9011 & 250 &  45.2 &  27.9 &  86.3 &  75.7 &  58.6  \\
HD\,20782   &      -     &  900 &     -     & 200 &  33.5 &  19.9 &  72.2 &  59.5 &  49.3  \\
HD\,23079   &      -     &  900 & 84.A-9004 & 200 &  34.1 &  19.7 &  80.8 &  68.5 &  51.8  \\
HD\,28185   &      -     &  500 & 83.A-9011 & 250 &  44.5 &  26.9 &  71.7 &  62.7 &  46.2  \\
HD\,30177   &      -     &  250 & 84.A-9004 & 200 &  48.5 &  31.0 &  72.2 &  62.0 &  53.3  \\
HD\,52265   &      -     &  250 & 80.A-9021 & 350 &  59.1 &  38.9 & 109.2 &  95.6 &  83.1  \\
HD\,65216   &      -     &  500 & 83.A-9003 & 250 &  22.2 &  12.4 &  54.4 &  50.3 &  35.7  \\
HD\,69830   &      -     &  800 & 77.C-0573 & 300 &  23.1 &  12.7 &  42.0 &  37.2 &  28.1  \\
HD\,73256   &      -     & 1100 & 83.A-9003 & 200 &  39.7 &  22.5 &  61.5 &  54.5 &  41.3  \\
HD\,75289   &      -     &  650 & 84.A-9003 & 500 &  57.3 &  36.4 & 108.2 &  95.4 &  78.3  \\
HD\,82943   &      -     &  700 & 84.A-9004 & 300 &  55.0 &  35.4 & 100.2 &  87.2 &  72.8  \\
HD\,92788   &      -     &  400 & 80.A-9021 & 250 &  49.0 &  30.0 &  81.1 &  70.1 &  58.3  \\
HD\,108147  &      -     &  650 & 83.A-9013 & 250 &  55.6 &  35.0 & 114.9 &  98.1 &  80.9  \\
HD\,111232  &      -     &  550 &     -     & 250 &  19.6 &  10.5 &  54.8 &  45.5 &  33.0  \\
HD\,114729  &      -     &  900 &     -     & 350 &  31.6 &  18.2 &  76.0 &  64.9 &  52.6  \\
HD\,117618  &      -     &  600 &     -     & 300 &  44.8 &  28.1 &  92.1 &  79.4 &  60.2  \\
HD\,134987  &      -     &  550 &     -     & 250 &  53.7 &  33.8 &  87.5 &  76.4 &  57.8  \\
HD\,160691  &  73.D-0578 &  600 &     -     & 250 &  54.1 &  35.1 &  89.2 &  77.1 &  65.6  \\
HD\,168443  &  72.C-0488 &  600 & 83.A-9003 & 300 &  41.6 &  25.5 &  72.9 &  65.5 &  50.7  \\
HD\,169830  &      -     &  550 & 83.A-9013 & 350 &  70.5 &  46.9 & 139.3 & 123.8 & 102.5  \\
HD\,179949  &      -     &  450 & 85.C-0743 & 450 &  59.7 &  38.2 & 118.5 & 101.9 &  83.4  \\
HD\,183263  &  75.C-0332 &  500 & 79.A-9013 & 250 &  57.4 &  37.5 &  98.7 &  84.4 &  67.7  \\
HD\,196050  &  72.C-0488 &  950 & 83.A-9011 & 300 &  56.8 &  37.1 &  97.5 &  82.6 &  69.5  \\
HD\,202206  &      -     &  700 &     -     & 200 &  43.3 &  26.7 &  74.7 &  67.0 &  48.7  \\
HD\,210277  &      -     &  900 &     -     & 200 &  41.1 &  25.0 &  65.1 &  57.7 &  44.4  \\
HD\,212301  &  82.C-0312 &  350 & 85.C-0743 & 350 &  58.4 &  37.2 & 112.8 &  97.6 &  81.8  \\
HD\,213240  &  72.C-0488 &  300 & 83.A-9011 & 200 &  51.9 &  33.5 &  99.1 &  82.5 &  71.1  \\
HD\,216435  &      -     &  700 &     -     & 200 &  61.2 &  40.1 & 108.9 &  96.5 &  77.1  \\
HD\,216437  &  80.D-0408 &  700 &     -     & 200 &  36.7 &  21.6 &  52.4 &  47.3 &  37.3  \\
HD\,216770  &  72.C-0488 &  600 &     -     & 120 &  54.9 &  36.9 &  96.3 &  81.8 &  65.7  \\
Ceres       &  60.A-9036 &  400 & 85.A-9027 & 350 &  36.3 &  21.4 &  72.3 &  61.2 &  48.2  \\
Ganymede    &      -     &  450 & 77.C-0766 & 250 &  35.6 &  21.6 &  71.3 &  62.4 &  49.0  \\
\noalign{\smallskip}
\hline
\end{tabular}
\end{table*}
}

\section{Analysis}
\label{sect:analysis}

\subsection{Model atmospheres and line broadening}
\label{sect:models}

For each star, a plane parallel (1D) model atmosphere was  
obtained from the standard
MARCS grid (Gustafsson et al. \cite{gustafsson08}) by interpolating
to the \teff , \logg , and \feh\ values of the star, and
the Uppsala program EQWIDTH was used to calculate
equivalent widths of the \CI\ and \OI\ lines
as a function of C or O abundance assuming LTE.
Interpolation to the observed equivalent widths then yields the LTE abundances.

Line data used in the analysis are given in Table \ref{table:linedata}.
The $gf$ values of the \CI\ lines are based on quantum mechanical calculations by
Hibbert et al. (\cite{hibbert93}) and those of oxygen are taken from 
Hibbert et al. (\cite{hibbert91}). Doppler broadening by non-thermal, small-scale
motions in the atmosphere is described by one parameter, $\xi_{\rm turb}$, in
the usual way. For the \OI\ lines, collisional broadening caused by neutral hydrogen
and helium atoms is based on the quantum mechanical calculations of 
Barklem et al. (\cite{barklem00}). Their work does not include the two \CI\ lines, 
for which the Uns{\"o}ld (\cite{unsold55}) approximation with an enhancement
factor of two was adopted as in Asplund et al. (\cite{asplund05}). 
The \CI\ lines are 
sufficiently weak to be insensitive to the assumed
enhancement factor; if it is changed from two to one, the derived [C/H] values change
by less than 0.01\,dex.

\begin{table}
\caption[ ]{Line data and derived solar abundances.}
\label{table:linedata}
\setlength{\tabcolsep}{0.10cm}
\begin{tabular}{cccrcccc}
\noalign{\smallskip}
\hline\hline
\noalign{\smallskip}
  ID & Wavelength & $\chi_{\rm exc}$ & log($gf$) & $EW_{\odot}$ &
 $A({\rm X})_{\odot}$\tablefootmark{a} &  $A({\rm X})_{\odot}$  \\
          &  (\AA )    &  (eV)  &   &  (m\AA )  & LTE  & non-LTE \\
\noalign{\smallskip}
\hline
\noalign{\smallskip}
\CI\ & 5052.17 & 7.685 & $-1.301$ & 35.9 & 8.44 & 8.43 \\
\CI\ & 5380.34 & 7.685 & $-1.616$ & 21.5 & 8.44 & 8.43 \\
\OI\ & 7771.94 & 9.146 &  0.369   & 71.8 & 8.88 & 8.66 \\
\OI\ & 7774.17 & 9.146 &  0.223   & 61.8 & 8.86 & 8.66 \\
\OI\ & 7775.39 & 9.146 &  0.002   & 48.6 & 8.84 & 8.67 \\
\noalign{\smallskip}
\hline
\end{tabular}
\tablefoot{
\tablefoottext{a}{For an element X,  $A({\rm X}) \equiv {\rm log} \, (N_{\rm X}/N_{\rm H}) +12.0$} 
}

\end{table}

\subsection{Non-LTE corrections}
\label{sect:non-LTE}

It is well know that the strength of the \OI\ triplet is subject to
large non-LTE effects (e.g. Kiselman \cite{kiselman01}). The most advanced
study is due to Fabbian et al. (\cite{fabbian09}), who calculated non-LTE 
corrections for a grid of MARCS models ranging from 4500 to 6500\,K in \teff ,
from 2.0 to 5.0 in \logg , and from $-3.0$ to 0.0 in \feh . 
Their model atom contains 54 energy levels and includes
electron collision cross sections based on quantum mechanical calculations by 
Barklem (\cite{barklem07}). An IDL program, made available, allows interpolation
of the non-LTE correction to the stellar values of \teff ,  \logg , and \feh\
for a given LTE oxygen abundance. For stars with $\feh > 0.0$, one needs
to make an extrapolation in \feh , but because the correction  changes
only a little from $\feh = -0.5$ to 0.0 (see Fig. 7  in Fabbian et al.), 
this is not expected to introduce any significant errors. 

The ``dark horse" in the non-LTE calculations is the handling of inelastic
collisions with hydrogen atoms, because no experimental or quantum mechanical
estimates of cross sections are available. Therefore, Fabbian et al. adopted
the classical Drawin formula (Drawin \cite{drawin68}) scaled by an empirical
factor $S_{\rm H}$. According to the 3D, non-LTE study of the solar centre-to-limb
variation of the \OI\ triplet lines by Pereira et al. (\cite{pereira09}), who
applied the same atomic model as Fabbian et al.,  $S_{\rm H} = 0.85$ leads to the
best agreement with the observations. Because Fabbian et al.  (\cite{fabbian09})
provide non-LTE corrections for both $S_{\rm H} = 0$ (H collisions are neglected)
and  $S_{\rm H} = 1$ (Drawin's formula), it is possible to interpolate their
corrections to $S_{\rm H} = 0.85$, which is the value applied in the present paper.

According to Table  \ref{table:linedata}, the solar non-LTE corrections range 
from $-0.22$ to $-0.17$ for the three \OI\ lines, 
i.e. close to $-0.20$\,dex on average.
As seen from Figs. 7 and 8 in  Fabbian et al.  (\cite{fabbian09}), the corrections
become increasingly more negative for higher \teff\ and lower \logg , whereas
stars with lower \teff\ than the Sun have less negative
non-LTE corrections. Altogether, this leads to significant differential
non-LTE corrections of \oh\ ranging from about $-0.2$\,dex at 
$\teff \simeq 6400$\,K to +0.1\,dex at $\teff \simeq 5400$\,K.

The non-LTE corrections for the \CI\ $\lambda 5052$ and $\lambda 5380$ lines
are much smaller than those for the \OI\ triplet. According to the detailed
calculations of Takeda \& Honda (\cite{takeda05}) adopted in this paper,
the corrections are 
$-0.01$\,dex for the Sun and changes only slightly as a function of \teff\ and
\logg . Their calculations were made on the assumption that
$S_{\rm H} = 1$. Assuming instead $S_{\rm H} = 0$,  Asplund et al.
(\cite{asplund05}) find a solar non-LTE correction of $-0.03$\,dex.
In this case the corrections of \ch\ would be somewhat higher,
but still close to be negligible compared to other error sources.

\subsection{Random and systematical errors}
\label{sect:errors}
Random errors arise from the uncertainties of the measured equivalent
widths and from the uncertainty of the adopted stellar atmospheric
parameters. The first source is best
estimated by comparing abundances obtained from different spectral lines
of the same element. The \ch\ abundances derived from the
two \CI\ lines agree very well; the average standard deviation of the mean 
is $\sigma$\ch\ = 0.009\,dex.  From the three \OI\ lines one gets  
$\sigma$\oh (non-LTE) = 0.017\,dex\footnote{Interestingly, this number increases
to 0.029\,dex for the LTE values of \oh\ showing the importance of individual 
non-LTE corrections for the three \OI\ lines.}. Adding these errors in quadrature,
the one-sigma error of \co\ due to $EW$ uncertainties becomes $\simeq 0.02$\,dex. 
 
\begin{figure}
\resizebox{\hsize}{!}{\includegraphics{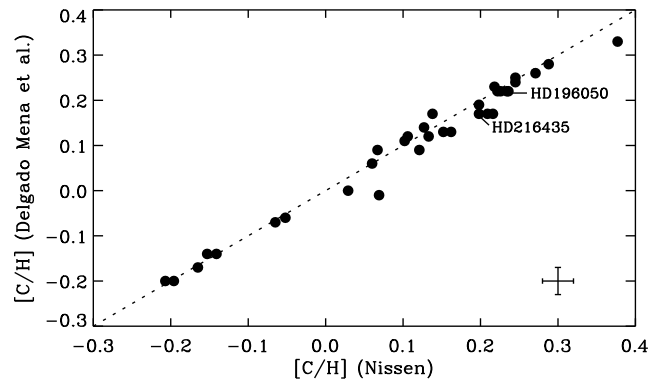}}
\caption{Carbon abundances of Delgado Mena et al. (\cite {delgado10})
in comparison with carbon abundances derived in this paper.}
\label{fig:CvsC}
\end{figure}

\begin{figure}
\resizebox{\hsize}{!}{\includegraphics{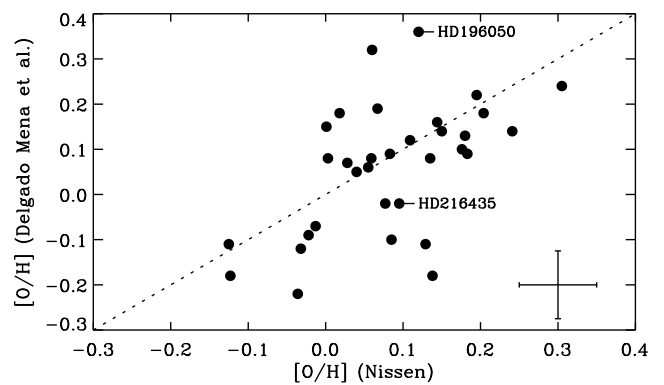}}
\caption{Oxygen abundances of Delgado Mena et al. (\cite {delgado10})
in comparison with non-LTE oxygen abundances derived in this paper.}
\label{fig:OvsO}
\end{figure}

\begin{table}
\caption[ ]{Changes in the derived non-LTE values\tablefootmark{a} of
\ch , \oh , and \co\ due to the
listed changes in  \teff , \logg , \feh , and \turb .}
\label{table:changes}
\setlength{\tabcolsep}{0.30cm}
\begin{tabular}{lccc}
\noalign{\smallskip}
\hline\hline
\noalign{\smallskip}
  Source & $\Delta \ch$ & $\Delta \oh$ & $\Delta \co$ \\
\noalign{\smallskip}
\hline
\noalign{\smallskip}

 $\Delta \teff = +45$\,K  &  $-0.030$  & $-0.059$ & $+0.029$ \\
 $\Delta \logg = +0.08$    &  $+0.026$  & $+0.034$ & $-0.008$ \\
 $\Delta \feh  = +0.04$    &  $+0.000$  & $+0.001$ & $-0.001$ \\
 $\Delta \turb  = +0.07$\,\kmprs   &  $-0.002$  & $-0.005$ & $+0.003$ \\
\noalign{\smallskip}
\hline
\end{tabular}
\tablefoot{
\tablefoottext{a}{Calculated for a representative star, \object{HD\,134987}}
}
\end{table}

Based on the rms deviations between spectroscopic and photometric parameters
discussed in Sect. \ref{sect:phot}, the one-sigma errors
for each set of parameters
are estimated to be $\sigma (\teff ) = 45$\,K, $\sigma (\logg ) = 0.08$,
$\sigma \feh  = 0.04$, and $\sigma (\turb )  = 0.07$\,\kmprs . The effects
on the derived \ch , \oh , and \co\ values of the corresponding
changes in stellar parameters are shown in Table \ref{table:changes}.
As seen, the largest contribution to the abundance errors 
arises from the error in \teff\ with \oh\ being more affected than \ch .
This is due to a higher excitation potential of the \OI\ lines 
(9.15\,eV) than in the case of the \CI\ lines (7.69\,eV), and to the
high \teff -sensitivity of the \OI\ non-LTE corrections.
Still, there is some canceling effect of the \teff\ dependence
for the \CI\ and the \OI\ lines; the error in \co\ is only half
the error in \oh . Furthermore, the derived \ch\ and \oh\ values have
nearly the same gravity dependence, so that the \logg -induced error
on \co\ becomes relatively small. Finally, it is noted that errors
induced by the uncertainties in \feh\ and \turb\ are negligible.

When adding all errors in quadrature, the total errors
become $\sigma \ch = 0.04$, $\sigma \oh = 0.07$, and $\sigma \co = 0.04$.   
These are the random one-sigma errors, but systematical errors may also
be present. As discussed in  Sect. \ref{sect:phot}, there may be
a systematic error of \teff\ as a function of \feh\ 
ranging from +60\,K at $\feh = -0.5$ to $-60$\,K at $\feh = +0.3$.
The corresponding change in \co\ ranges from +0.04 at
$\feh = -0.5$ to $-0.04$ at $\feh = +0.3$. There may also be an 
effect of applying 1D model atmospheres instead of more realistic
3D models. According to Asplund et al. (\cite{asplund04}),
the difference of the oxygen abundance derived from the
\OI\ triplet using a 3D and a MARCS model for the solar atmosphere
is $-0.06$\,dex, whereas the corresponding difference for
the solar carbon abundance determined from the \CI\
lines is $-0.01$\,dex (Asplund et al. \cite{asplund05}). 
It remains to be seen if these 3D -- MARCS differences  
change with \teff\ or \feh , but because the range of parameters for 
the present sample of stars
is less than $\pm 600$\,K in \teff\ and $\pm 0.5$\,dex in
\feh , I do not expect any major 3D effects on the derived \co\ values.
It is also noted that the uncertainty of the $S_{\rm H}$ parameter
for hydrogen collisions, say a change from the adopted value $S_{\rm H} = 0.85$
to an extreme of $S_{\rm H} = 0$, has only a small effect on \co ;
the changes are within $\pm 0.02$\,dex.

\section{Results based on spectroscopic stellar parameters}
\label{sect:spec}

Carbon and oxygen abundances are first derived with
MARCS model atmospheres having the same parameters as
applied by Delgado Mena et al. (\cite{delgado10}). The original
source of these parameters is  Sousa et al. (\cite{sousa08}), who in
a differential, LTE analysis with respect to the Sun determined 
\teff , \logg , \feh , and \turb\ by requesting that  
iron abundances derived from \FeI\ and \FeII\ lines have no
systematic dependence on excitation potential, ionization stage, or
equivalent width.
The parameters are listed in Table \ref{table:results} together with
the LTE and non-LTE values of \ch\ and \oh\ derived in this paper. 

\onltab{4}{
\begin{table*}
\caption[ ]{Adopted atmospheric parameters and derived carbon and oxygen abundances}
\label{table:results}
\setlength{\tabcolsep}{0.30cm}
\begin{tabular}{lccrcrrrrr}
\noalign{\smallskip}
\hline\hline
\noalign{\smallskip}
  ID & \teff & \logg & \feh & \turb & \ch &   \ch & \oh &   \oh & C/O  \\
     &  (K)  &       &      &\kmprs & LTE & non-LTE & LTE & non-LTE & non-LTE  \\
\noalign{\smallskip}
\hline
\noalign{\smallskip}
HD\,142 &    6403 &   4.62 &   0.09 &   1.74 &   0.21 &   0.21 &   0.21 &   0.08 &   0.78  \\
HD\,1237 &    5514 &   4.50 &   0.07 &   1.09 &   0.07 &   0.07 &   0.07 &   0.14 &   0.50  \\
HD\,4308 &    5644 &   4.38 &  $-$0.34 &   0.90 &  $-$0.15 &  $-$0.15 &   0.04 &   0.09 &   0.34  \\
HD\,16141 &    5806 &   4.19 &   0.16 &   1.11 &   0.07 &   0.06 &   0.12 &   0.06 &   0.59  \\
HD\,20782 &    5774 &   4.37 &  $-$0.06 &   1.00 &  $-$0.06 &  $-$0.07 &  $-$0.03 &  $-$0.03 &   0.54  \\
HD\,23079 &    5980 &   4.48 &  $-$0.12 &   1.12 &  $-$0.14 &  $-$0.14 &  $-$0.09 &  $-$0.12 &   0.55  \\
HD\,28185 &    5667 &   4.42 &   0.21 &   0.94 &   0.22 &   0.22 &   0.13 &   0.14 &   0.71  \\
HD\,30177 &    5588 &   4.29 &   0.38 &   1.08 &   0.38 &   0.38 &   0.31 &   0.31 &   0.68  \\
HD\,52265 &    6136 &   4.36 &   0.21 &   1.32 &   0.15 &   0.15 &   0.18 &   0.05 &   0.72  \\
HD\,65216 &    5612 &   4.44 &  $-$0.17 &   0.78 &  $-$0.20 &  $-$0.19 &  $-$0.09 &  $-$0.04 &   0.40  \\
HD\,69830 &    5402 &   4.40 &  $-$0.06 &   0.80 &  $-$0.05 &  $-$0.05 &  $-$0.10 &  $-$0.02 &   0.55  \\
HD\,73256 &    5526 &   4.42 &   0.23 &   1.11 &   0.22 &   0.22 &   0.13 &   0.18 &   0.64  \\
HD\,75289 &    6161 &   4.37 &   0.30 &   1.29 &   0.11 &   0.10 &   0.14 &   0.00 &   0.73  \\
HD\,82943 &    5989 &   4.43 &   0.26 &   1.10 &   0.20 &   0.20 &   0.22 &   0.14 &   0.66  \\
HD\,92788 &    5744 &   4.39 &   0.27 &   0.95 &   0.24 &   0.24 &   0.21 &   0.18 &   0.66  \\
HD\,108147 &    6260 &   4.47 &   0.18 &   1.30 &   0.06 &   0.06 &   0.13 &   0.00 &   0.66  \\
HD\,111232 &    5460 &   4.43 &  $-$0.43 &   0.62 &  $-$0.17 &  $-$0.16 &   0.04 &   0.13 &   0.30  \\
HD\,114729 &    5844 &   4.19 &  $-$0.28 &   1.23 &  $-$0.21 &  $-$0.21 &  $-$0.09 &  $-$0.12 &   0.47  \\
HD\,117618 &    5990 &   4.41 &   0.03 &   1.13 &   0.03 &   0.03 &   0.05 &  $-$0.01 &   0.63  \\
HD\,134987 &    5740 &   4.30 &   0.25 &   1.08 &   0.29 &   0.29 &   0.24 &   0.20 &   0.70  \\
HD\,160691 &    5780 &   4.27 &   0.30 &   1.09 &   0.27 &   0.27 &   0.25 &   0.19 &   0.69  \\
HD\,168443 &    5617 &   4.22 &   0.06 &   1.21 &   0.14 &   0.14 &   0.15 &   0.15 &   0.56  \\
HD\,169830 &    6361 &   4.21 &   0.18 &   1.56 &   0.13 &   0.12 &   0.29 &   0.07 &   0.66  \\
HD\,179949 &    6287 &   4.54 &   0.21 &   1.36 &   0.13 &   0.12 &   0.16 &   0.04 &   0.70  \\
HD\,183263 &    5991 &   4.38 &   0.34 &   1.23 &   0.22 &   0.22 &   0.15 &   0.06 &   0.83  \\
HD\,196050 &    5917 &   4.32 &   0.23 &   1.21 &   0.23 &   0.23 &   0.20 &   0.12 &   0.75  \\
HD\,202206 &    5757 &   4.47 &   0.29 &   1.01 &   0.16 &   0.16 &   0.10 &   0.08 &   0.70  \\
HD\,210277 &    5505 &   4.30 &   0.18 &   0.86 &   0.25 &   0.25 &   0.20 &   0.24 &   0.59  \\
HD\,212301 &    6271 &   4.55 &   0.18 &   1.29 &   0.12 &   0.12 &   0.13 &   0.02 &   0.73  \\
HD\,213240 &    5982 &   4.27 &   0.14 &   1.25 &   0.10 &   0.10 &   0.12 &   0.03 &   0.68  \\
HD\,216435 &    6008 &   4.20 &   0.24 &   1.34 &   0.20 &   0.19 &   0.22 &   0.09 &   0.73  \\
HD\,216437 &    5887 &   4.30 &   0.25 &   1.31 &   0.23 &   0.22 &   0.18 &   0.11 &   0.75  \\
HD\,216770 &    5424 &   4.38 &   0.24 &   0.91 &   0.25 &   0.25 &   0.12 &   0.18 &   0.68  \\
Sun        &    5777 &   4.44 &   0.00 &   1.00 &   0.00 &   0.00 &   0.00 &   0.00 &   0.58  \\
\noalign{\smallskip}
\hline
\end{tabular}

\end{table*}
} 

As seen from Fig. \ref{fig:CvsC}, there is an excellent agreement between
the carbon abundances derived by  Delgado Mena et al. (\cite {delgado10})
and the values derived in this paper. The rms deviation in \ch\ is only 0.025\,dex.
This is not surprising, because in both studies C abundances are
derived from equivalent widths of the $\lambda \lambda 5052, 5380$ \CI\ lines
in high $S/N$ HARPS spectra using the same set of stellar parameters.

\begin{figure}
\resizebox{\hsize}{!}{\includegraphics{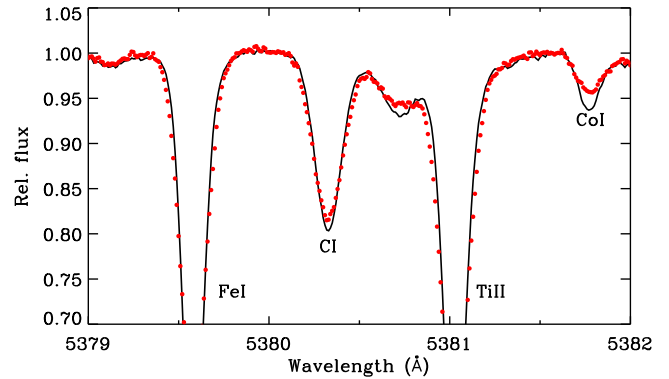}}
\caption{The $\lambda 5380$ \CI\ line in HARPS spectra of
\object{HD\,196050} (full drawn line) and \object{HD\,216435} (red dots).}
\label{fig:CI5380}
\end{figure}

\begin{figure}
\resizebox{\hsize}{!}{\includegraphics{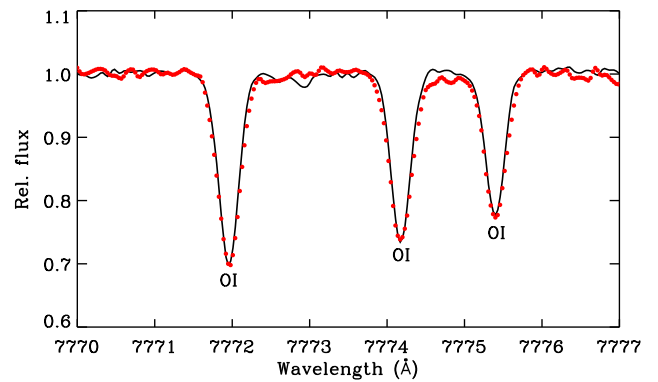}}
\caption{The \OI\ triplet in FEROS spectra of
\object{HD\,196050} (full drawn line) and \object{HD\,216435} (red dots).}
\label{fig:OItriplet}
\end{figure}

\begin{figure}
\resizebox{\hsize}{!}{\includegraphics{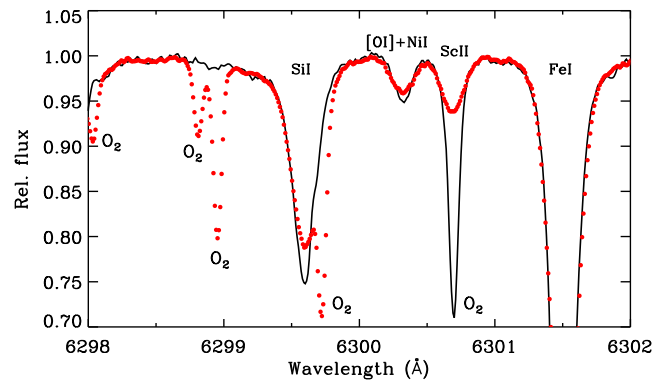}}
\caption{The $\lambda 6300.3$ [\OI ]+\NiI\ blend in HARPS spectra of
\object{HD\,196050} (full drawn line) and \object{HD\,216435} (red dots).}
\label{fig:OI6300}
\end{figure}

The situation is, however, very different when one compares the oxygen abundances
derived by  Delgado Mena et al. (\cite {delgado10}) from the $\lambda 6300$
[\OI ] line with O abundances derived in this paper from FEROS spectra of
the $\lambda 7774$ \OI\ triplet (see Fig. \ref{fig:OvsO}). In this case, the 
rms deviation is as large as 0.12\,dex.
The problem can be illustrated by comparing the spectra of two stars
with similar atmospheric parameters \teff , \logg ,  and \feh , i.e \object{HD\,196050}
(5917\,K, 4.32, 0.23) and \object{HD\,216435} (6008\,K, 4.20, 0.24). As seen
from Figs. \ref{fig:CI5380} and \ref{fig:OItriplet}, the \CI\ and \OI\ lines
have about the same strengths in the two stars. Actually, the lines of 
\object{HD\,216435} are a bit wider and have slightly larger $EW$s 
than those of \object{HD\,196050}, but due to the
small differences in \teff\ and \logg\ between the two stars,
this is to be expected for fixed values of \ch\ and \oh .
As seen from Table \ref{table:results}, the model-atmosphere
analysis provide very similar O abundances,
i.e.  $\oh = 0.12$ for \object{HD\,196050} and $\oh = 0.09$ in the case of 
\object{HD\,216435}. This is in stark contrast to the results of 
Delgado Mena et al. (\cite {delgado10}), who obtain $\oh = 0.36$ for
\object{HD\,196050} and $\oh = -0.02$ for \object{HD\,216435}.

Figure \ref{fig:OI6300} shows a comparison of HARPS spectra of 
\object{HD\,196050} and \object{HD\,216435} in the 6300\,\AA\
region. The [\OI ]+\NiI\ blend is affected by a telluric O$_2$ line 
in many of the available HARPS spectra of \object{HD\,196050},
so the spectrum shown in Fig. \ref{fig:OI6300} is an average
of a subset of spectra from July 2004, when the topocentric radial 
velocity of \object{HD\,196050} shifts the telluric line away from the [\OI ] line.
As seen from the figure the equivalent width of the [\OI ]+\NiI\ blend is nearly
the same in the spectra of \object{HD\,196050} and \object{HD\,216435}, i.e.
$EW$ = 8.0 and 7.5\,m\AA , respectively. Adopting 
[Ni/H] = 0.28 and 0.29 for the two stars (as used by
Delgado Mena et al.) and solar flux data $EW$[\OI ] = 3.5\,m\AA\
and $EW$(\NiI ) = 2.0\,m\AA\ (Caffau et al. \cite{caffau08}),
a differential analysis with respect to the Sun
leads to $\oh = 0.17 \pm 0.10$ for \object{HD\,196050}
and $\oh = 0.11 \pm 0.10$ for \object{HD\,216435}, where the main
contribution to the error arises from the uncertainty of 
setting the continuum level. These \oh\ values agree
well with the non-LTE values derived from the \OI\ triplet, 
but lend no support to the large difference in \oh\ found by
Delgado Mena et al. (\cite {delgado10}). Perhaps, they applied a different set of HARPS  
spectra for \object{HD\,196050}; in any case, the analysis shows that
oxygen abundances derived from the [\OI ]+\NiI\ blend in spectra of metal-rich
stars are uncertain due to difficulties in setting the continuum
level and correcting for the large contribution of the \NiI\ line.

\section{Results based on photometric stellar parameters}
\label{sect:phot}

In order to test the reliability of the derived C/O ratios,
photometric atmospheric parameters of the stars have been determined 
as an alternative to the spectroscopic parameters 
adopted by  Delgado Mena et al. (\cite {delgado10}). \teff\ was taken
as the mean value 
derived from the \by\ and \VK\ colour indices using the IRFM calibrations
of Casagrande et al. ({\cite{casagrande10}) with $V$ magnitudes and
\by\ from Olsen (\cite{olsen83}) and $K$ magnitudes from the 2MASS catalogue
(Skrutskie et al. \cite{skrutskie06}). All stars are closer than 60\,pc
according to their Hipparcos parallaxes, so reddening corrections can be neglected. 

The surface gravities of the stars were determined from the fundamental relation
\begin{eqnarray} \log \frac{g}{g_{\odot}}  =  \log \frac{\cal{M}}{\cal{M}_{\odot}} +
4 \log \frac{\teff}{T_{\rm eff,\odot}} + 0.4 (M_{\rm bol} - M_{{\rm bol},\odot})
\end{eqnarray} where $\cal{M}$ is the mass of the star and $M_{\rm bol}$ the absolute
bolometric magnitude. Hipparcos parallaxes (van Leeuwen \cite{leeuwen07})  were
used to derive \Mv\ and the bolometric correction adopted from
Casagrande et al. ({\cite{casagrande10}).
The stellar mass was obtained by interpolating in the luminosity - \logteff\
diagram between the Yonsei -Yale
evolutionary tracks by Yi et al. (\cite{yi03}) as described in 
detail by Nissen \& Schuster (\cite{nissen12}). 
Due to the small error of the Hipparcos parallaxes,
the estimated error of \logg\ is only 0.04\,dex.

The metallicity was determined from 12 \FeII\ lines
with equivalent widths being measured from the HARPS spectra.
By using \FeII\ lines only, the derived
\feh\ value is unlikely to be affected by departures from LTE
(Mashonkina et al. \cite{mashonkina11}, Lind et al. \cite{lind12}). 
The \FeII\ lines have 
equivalent widths in the range $10 \simlt EW \simlt 90$\,m\AA ,
which allows a determination of \turb\ from the requirement
that the derived \feh\ should not depend on $EW$.

\begin{figure}
\resizebox{\hsize}{!}{\includegraphics{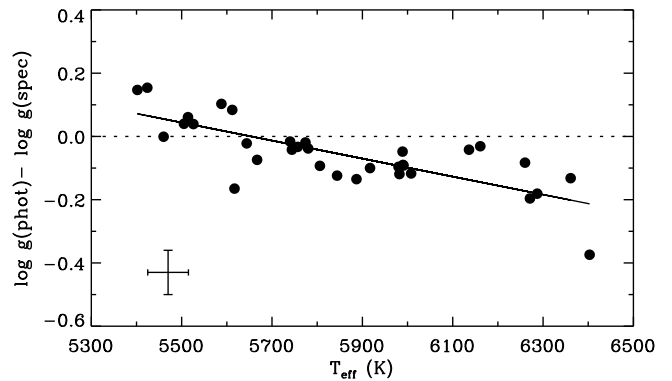}}
\caption{Differences between photometric gravities determined via Hipparcos
parallaxes and spectroscopic gravities used by
Delgado Mena et al. (\cite{delgado10})}
\label{fig:delta.logg}
\end{figure}

\begin{figure}
\resizebox{\hsize}{!}{\includegraphics{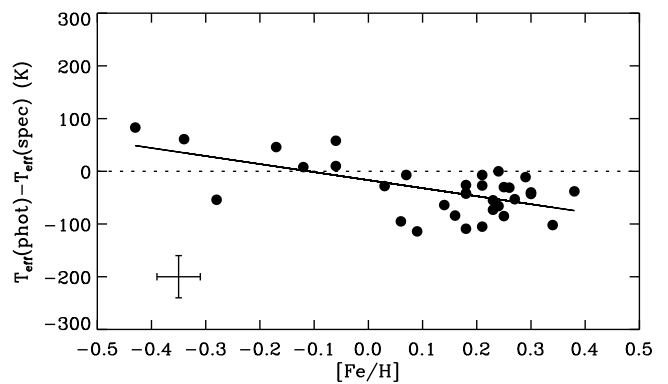}}
\caption{Differences between photometric values of \teff\
determined from colour indices and
spectroscopic values used by Delgado Mena et al. (\cite{delgado10}).}
\label{fig:delta.Teff}
\end{figure}

\begin{figure}
\resizebox{\hsize}{!}{\includegraphics{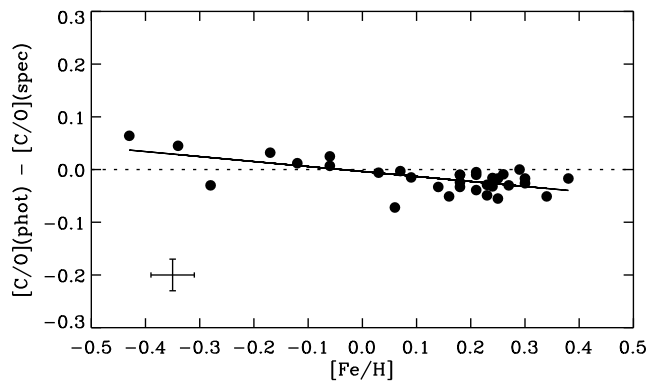}}
\caption{Difference between \co\ values derived in this paper,
when using photometric and spectroscopic stellar parameters, respectively.}
\label{fig:delta.CO}
\end{figure}

It is noted that this procedure for the determination of stellar parameters
has to be iterated, because the \teff\ calibrations, bolometric corrections
and mass determinations depend on \feh .

When comparing the photometric and spectroscopic parameters,
the following rms deviations are obtained:
rms$(\Delta \teff )$ = 62\,K, rms$(\Delta \logg )$ = 0.12,
rms$(\Delta \feh )$ = 0.05, and rms$(\Delta \turb )$ = 0.10\,\kmprs .
The scatter for \feh\ and \turb\ is as small as one would expect,
if the one-sigma errors of each of the two studies are 
$\sigma \feh \simeq 0.03$ and $\sigma (\turb ) = 0.07$\,\kmprs .   
In the case of \logg , the scatter is larger than expected
from the estimated errors, i.e. $\sigma (\logg ) = 0.06$ for the spectroscopic
gravities  (Sousa et al. \cite{sousa08}) and $\sigma (\logg ) = 0.04$
for the photometric values obtained via the Hipparcos parallaxes.
As shown in Fig. \ref{fig:delta.logg}, this is mainly due to
a significant dependence of $\Delta \logg $ on \teff ; the rms scatter
around the fitted line is only 0.07\,dex. This
trend is not easy to explain; according to  Lind et al. ( \cite{lind12}, Fig. 4)
non-LTE effects on the ionization balance of Fe (used to determine
the spectroscopic gravities) are almost negligible for the effective temperatures,
gravities and metallicities of the present sample of stars. 

A similar problem is present, when one compares photometric and
spectroscopic temperatures. In this case, the difference depends on
\feh\ as shown in Fig. \ref{fig:delta.Teff}. Here, the rms deviation
around the fit is 40\,K. Again it is unclear, what is causing
this trend with \feh .

Despite of these systematic differences between the photometric and
spectroscopic stellar parameters, the C/O ratios derived in this paper
are not much affected as seen from Fig. \ref{fig:delta.CO}.
[C/O] is immune to changes in \logg , whereas the trend of $\Delta \teff $
with \feh\ shown in Fig. \ref{fig:delta.Teff}
introduces a small dependence of $\Delta \co$ on \feh .

\section{Discussion}
\label{sect:discussion}

Fig. \ref{fig:COratios} shows the C/O ratios derived in this paper
compared to
the ratios derived by  Delgado Mena et al. (\cite{delgado10})
with the same set of spectroscopic stellar parameters applied in the two studies.
Evidently, the C/O ratios of Delgado Mena et al. have a much larger scatter 
than the ratios obtained in this paper.
Furthermore, the  Delgado Mena et al.  C/O ratios
tend to be higher than the values in the present study,
which is mainly due to the fact that they adopt a solar ratio
(C/O)$_{\odot} = 0.66$ instead of the value (C/O)$_{\odot} = 0.58$
corresponding to the solar abundances derived in this paper
(see Table \ref{table:results}). Recent 3D studies lead to nearly the
same C/O ratio in the Sun, i.e (C/O)$_{\odot} = 0.55 \pm 0.10$  
(Asplund  et al. \cite{asplund09}; Caffau et al. \cite{caffau08}, \cite{caffau10}).
The solar C/O ratio adopted by Delgado Mena et al. is
on the high side of the quoted error range for these 3D studies. 

\begin{figure}
\resizebox{\hsize}{!}{\includegraphics{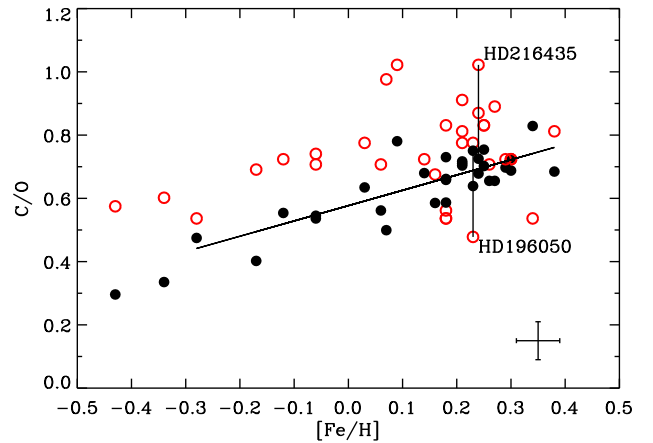}}
\caption{Carbon-to-oxygen ratios derived in this paper
(filled, black circles) in comparison with
the C/O ratios determined by  Delgado Mena et al. (\cite{delgado10})
(open, red circles). The error bars shown refer to data from this
paper. In the case of \object{HD\,196050} and \object{HD\,216435}, the 
filled and open circles are connected by a straight line.}
\label{fig:COratios}
\end{figure}

Excluding two stars with $\feh < -0.3$ having thick-disk kinematics,
one obtains a linear fit,
C/O = 0.58 + 0.48 \feh, to the data of this pape. If the two 
thick-disk are included the relation becomes C/O = 0.56 + 0.54 \feh. 
In both cases, the rms dispersion around the fit is
$\sigma$\,(C/O) = 0.06. In comparison, the Delgado Mena et al. (\cite{delgado10})
data have a scatter of $\sigma$ = 0.14 relative to a linear fit with 
three stars having C/O $\simeq 1$ and 10 stars having C/O $> 0.8$.
As discussed in Sect. \ref{sect:spec}, this
difference in the C/O distribution is mainly 
due to the large differences in the oxygen abundances derived from
the $\lambda 6300$ [\OI ] line by Delgado Mena et al.
and the O abundances derived in this paper from the \OI\ triplet
as exemplified by the comparison of \object{HD\,196050} and 
\object{HD\,215435}. The small dispersion of C/O derived 
in this paper suggests that the high C/O values ($> 0.8$)
found by  Delgado Mena et al. (\cite{delgado10}) and by
Petigura \& Marcy (\cite{petigura11}) are spurious due
to difficulties in determining oxygen abundances from the
$\lambda 6300$ [\OI ]+\NiI\ blend especially at high metallicities. 
In the solar spectrum, the $EW$s of the \NiI\ and [\OI ] lines
are 2.0 and 3.5\,m\AA , respectively, (Caffau et al. 
\cite{caffau08}) i.e. a ratio of $R \simeq  0.55$, but in a solar-type star
with $\feh = +0.3$, $\ofe = -0.15$, and $\nife = +0.1$
(Bensby et al. \cite{bensby05}), the ratio is increased to
$R \simeq 1$.

In view of the possible systematic errors in the 3D, non-LTE
corrections for the C and O abundances derived from 
high-excitation \CI\ and \OI\ lines, it is important to
check the trend and scatter of C/O obtained in this paper
by independent determinations of C/O. Such a 
study has been carried out by Bensby et al. 
(\cite{bensby04}) and Bensby \& Feltzing (\cite{bensby06}), 
who determined O and C
abundances from weak forbidden lines, i.e. [\OI ] at
6300\,\AA\ and [\CI ] at 8727\,\AA . The non-LTE corrections
for these lines are vanishing and 3D corrections are 
small in the solar case (Asplund et al. \cite{asplund04},
\cite{asplund05}; Caffau et al. \cite{caffau08}, \cite{caffau10}),
but the lines are weak and affected by blends; in addition to the
\NiI\ blend of the $\lambda 6300$ [\OI ] line, a weak \FeI\ line
is overlapping the $\lambda 8727$ [\CI ] line. Hence, very high
resolution and high $S/N$ spectra are needed like those
obtained with the Coud\'{e} Echelle Spectrograph
at the ESO 3.6\,m telescope by Bensby et al.  

Figure \ref{fig:CO-FeH} shows \co\ versus \feh\ for a sample
of 42 dwarf stars with $5000 \! < \teff < \! 6500$\,K from Table\,1
in Bensby \& Feltzing (\cite{bensby06}), divided into thin- and  
thick-disk stars according to their kinematics.
In addition, the figure shows data from the present study
based on model atmospheres with photometric parameters.
For this sample, three stars with a total space velocity 
relative to the Local Standard of Rest
$V_{\rm LSR} > 80$\,\kmprs (Holmberg et al. \cite{holmberg09}) 
have been classified as belonging to the thick disk. 
Four stars are in common between the two samples;
they show good agreement in [C/O], i.e. a mean  and rms deviation
(Bensby $-$ this paper) $\Delta \co  = 0.02 \pm 0.05$.

As seen from Fig. \ref{fig:CO-FeH}, there is an excellent agreement
between [C/O] data from  Bensby \& Feltzing (\cite{bensby06}) and
from this paper, including a systematic difference in \co\ between
thin- and thick-disk stars at low metallicity. A linear fit to
the thin-disk data yields a trend
\begin{eqnarray}
\co = -0.002 + 0.22 \,\,\feh ,
\end{eqnarray}
with a standard deviation of $\pm 0.048$\,dex for 30 stars from this work
and $\pm 0.068$\,dex for 27 stars from Bensby \& Feltzing.
This equation corresponds to $\co = 0.086$ at $\feh = +0.4$ or
C/O = 0.71, when adopting a solar ratio (C/O)$_{\odot} = 0.58$.
If the solar ratio is (C/O)$_{\odot} = 0.65$, Eq. (2) corresponds
to C/O = 0.79 at $\feh = +0.4$.

Another two high-precision studies of carbon and oxygen abundances 
in solar-type stars (Takeda \& Honda \cite{takeda05};
Ram\'{\i}rez et al. \cite{ramirez09}) also
show good agreement with the \co --\feh\ trend found 
in this paper as seen from Fig. \ref{fig:CO-FeH.Ramirez-Takeda}. 
In both works, the adopted abundances are obtained from
high-excitation \CI\ and \OI\ lines with non-LTE corrections
based on their own calculations.
The \co\ values from Takeda \& Honda, which include 27 planet-harbouring
stars, distribute around the line given in Eq. (2) with an rms 
deviation of $\pm 0.070$\,dex, and the data from Ram\'{\i}rez et al.
for solar twins and analogs have an rms deviation of 0.058\,dex.  
In both cases, the scatter can be explained by the errors of the 
abundance determination, and although a few stars from 
Takeda \& Honda  fall above $\co = 0.14$, which corresponds to
C/O = 0.8, this may be accidental.                       

\begin{figure}
\resizebox{\hsize}{!}{\includegraphics{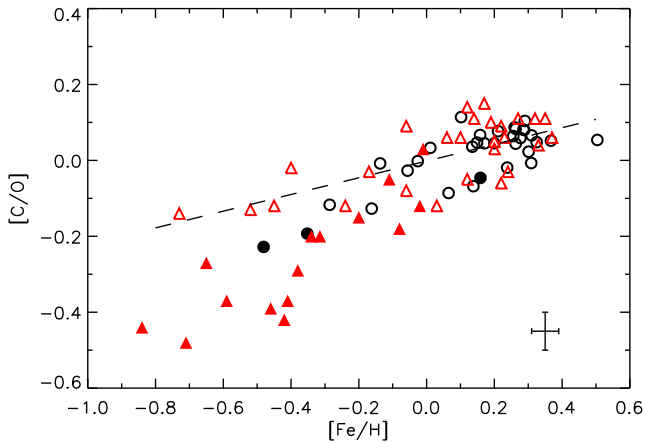}}
\caption{\co\ versus \feh . Data from this paper
are plotted with circles and data from Bensby \& Feltzing
(\cite{bensby06}) are shown with (red) triangles.
Open symbols refer to stars with thin-disk kinematics and
filled symbols to thick-disk stars. A linear fit to the thin-disk
stars, given in Eq. (2), is shown by a dashed line.}
\label{fig:CO-FeH}
\end{figure}

\begin{figure}
\resizebox{\hsize}{!}{\includegraphics{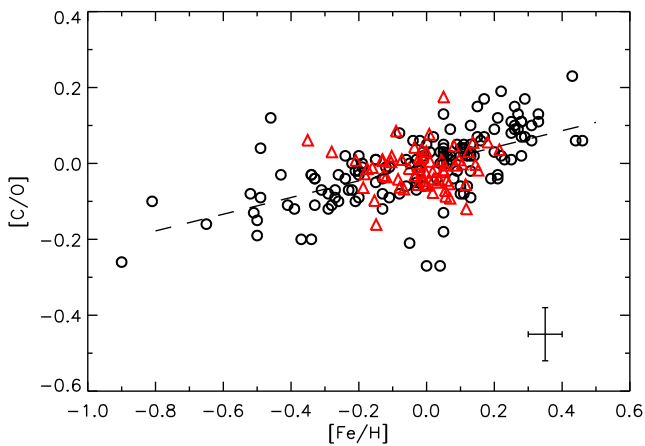}}
\caption{\co\ versus \feh\ with data from Takeda \& Honda
(\cite{takeda05}) (circles) and  Ram\'{\i}rez et al.
(\cite{ramirez09}) (red triangles). Eq. (2) is shown as
a dashed line.}
\label{fig:CO-FeH.Ramirez-Takeda}
\end{figure}

While oxygen is probably formed on a relatively short time-scale in
massive stars exploding as Type II SNe, the origin of carbon is more
uncertain with many possible sources including Type II supernovae, novae,
winds from massive stars, and low- to intermediate-mass stars 
(Gustafsson et al. \cite{gustafsson99}).
As shown by Akerman et al. (\cite{akerman04}),
\co\ raises from a level of $-0.5$\,dex in halo stars with $\oh < -1$
to $\co \simeq 0$ in disk stars with solar metallicity. A similar trend
is found from \HII\ regions in spiral and irregular Galaxies
(Garnett et al. \cite{garnett04}). From this
apparent universal metallicity dependence, Akerman et al. argue that
the carbon enrichment in the Galactic disk
is mainly due to metal-dependent winds from massive
stars. In order to explain the difference in \co\ between thin- and thick-disk
stars shown in Fig. \ref{fig:CO-FeH}, it is, however, necessary to invoke 
a significant carbon contribution from low- and intermediate-mass stars ocurring
on a relatively long time-scale in the Galactic thin disk 
(Chiappini et al. \cite{chiappini03}; Carigi et al. \cite{carigi05};
Cescutti et al. \cite{cescutti09}), 
whereas data for \co\ in stars belonging to the thick disk and the Galactic bulge may be
explained by metallicity-dependent C and O yields for massive stars alone
(Cescutti et al. \cite{cescutti09}). The situation is, however, complicated
and the \co\ data at high metallicity are not well explained, so
more precise observations of C/O, more reliable yields, and realistic
Galactic models are needed to advance the field. 

\section{Conclusions}
\label{sect:conclusions}

The results in this paper and those of 
Bensby \& Feltzing (\cite{bensby06}), Takeda \& Honda (\cite{takeda05}),
and  Ram\'{\i}rez et al. (\cite{ramirez09}) show a tight \co --\feh\ correlation
for thin-disk stars.
Furthermore, \co\ is not reaching higher than $\co \simeq 0.1$\,dex
(corresponding to C/O\,$\simeq 0.7$ for a solar ratio of (C/O)$_{\odot} = 0.55$) even at
the highest metallicities. This suggests that the recent findings 
by Delgado Mena et al. (\cite{delgado10}) and Petigura \& Marcy
(\cite{petigura11}) of C/O $> 0.8$ in a significant fraction (10 -- 15\,\%) of
solar-type stars are spurious due to difficulties in measuring the strength
of the weak $\lambda 6300$ [\OI ]+\NiI\ blend with sufficient precision and to
correct for the \NiI\ contribution when deriving oxygen abundances.

As mentioned in the introduction, modelling of planet formation 
(Bond et al. \cite{bond10}) indicates that C/O\,$> \! 0.8$
in proto-planetary disks with uniform composition
lead to the formation of terrestrial ``carbon planets"  
consisting of carbides and graphite instead of ``Earth-like" silicates.
Assuming that the composition of a proto-planetary disk is the same
as that of the host star, the results in this paper (C/O $< 0.8$) lend no support to
the existence of such carbon planets. 

The separation of \co\ between thin- and thick-disk stars suggests that
carbon in the thin disk are mainly formed in low- and
intermediate-mass stars. Assuming that oxygen comes from massive stars exploding
as Type II SNe, the small dispersion of \co\ indicates that the nucleosynthesis
products of these objects are well mixed in the interstellar medium
before new stars are formed.
A detailed understanding of the formation of the two elements requires,
however, further observational and theoretical studies.

\begin{acknowledgements}
The referee, Elisabetta Caffau, is thanked for comments that helped to
improve the manuscript significantly.
Funding for the Stellar Astrophysics Centre is provided by the
Danish National Research Foundation (Grant agreement no.: DNRF106).
The research is supported by the ASTERISK project
(ASTERoseismic Investigations with SONG and Kepler)
funded by the European Research Council (Grant agreement no.: 267864).
This publication made use of the SIMBAD database operated
at CDS, Strasbourg, France, and of data products from the Two Micron All
Sky Survey, which is a joint project of the University of Massachusetts and
the Infrared Processing and Analysis Center/California Institute of
Technology, funded by NASA and the National Science Foundation. 
\end{acknowledgements}

\Online


\begin{thebibliography}{}

\bibitem[2004]{akerman04}
Akerman, C. J., Carigi, L., Nissen, P. E., Pettini, M., \& Asplund, M. 
2004, A\&A, 414, 931


\bibitem[2004]{asplund04}
Asplund, M., Grevesse, N., Sauval, A. J., Allende Prieto, C., \&
Kiselman, D. 2004, A\&A, 417, 751

\bibitem[2005]{asplund05}
Asplund, M., Grevesse, N., Sauval, A. J., Allende Prieto, C., \&
Blomme, R. 2005, A\&A, 431, 693

\bibitem[2009]{asplund09}
Asplund, M., Grevesse, N., Sauval, A. J., \& Scott, P. 2009, ARA\&A, 47, 481

\bibitem[2007]{barklem07}
Barklem, P. S. 2007, A\&A, 462, 781

\bibitem[2000]{barklem00}
Barklem, P. S., Piskunov, N., \& O'Mara, B. J. 2000, A\&AS, 142, 467

\bibitem[2006]{bensby06}
Bensby, T., \& Feltzing, S. 2006, MNRAS, 367, 1181

\bibitem[2004]{bensby04}
Bensby, T., Feltzing, S., \& Lundstr{\"o}m, 2004, A\&A, 415, 155

\bibitem[2005]{bensby05}
Bensby, T., Feltzing, S., Lundstr{\"o}m, I., Ilyin, I. 2005, A\&A, 433, 185

\bibitem[2010]{bond10}
Bond, J. C., O'Brien, D. P., \& Lauretta, D. S. 2010, ApJ, 715, 1050

\bibitem[2010]{caffau10}
Caffau, E., Ludwig, H.-G., Bonifacio, P., et al. 2010,  A\&A, 514, A92

\bibitem[2008]{caffau08}
Caffau, E., Ludwig, H.-G., Steffen, M., et al. 2008,  A\&A, 488, 1031

\bibitem[2005]{carigi05}
Carigi, L., Peimbert, M., Esteban, C., \& Garc\'{\i}a-Rojas, J. 2005, ApJ, 623,  213

\bibitem[2010]{casagrande10}
Casagrande, L., Ram\'{\i}rez. I., Mel\'{e}ndez, J., Bessell, M., \& Asplund, M. 2010,
A\&A, 512, A54 

\bibitem[2009]{cescutti09}
Cescutti, G., Matteucci, F., McWilliam, A., \& Chiappini, C. 2009, A\&A, 505, 605

\bibitem[2003]{chiappini03}
Chiappini, C., Matteucci, F., \& Meynet, G. 2003, A\&A, 410, 257

\bibitem[2010]{delgado10}
Delgado Mena, E., Israelian, G., Gonz\'{a}lez Hern\'{a}ndez, J. I., et al. 2010,
ApJ, 725, 2349

\bibitem[2011]{demory11}
Demory, B.-O., Gillon, M., Deming, D., et al. 2011, A\&A, 533, A114

\bibitem[1968]{drawin68}
Drawin, H.-W. 1968, Zeitschrift f{\"u}r Physik, 211, 404

\bibitem[1997]{esa97}
ESA 1997, The Hipparcos and Tycho Catalogues, ESA SP-1200

\bibitem[2009]{fabbian09}
Fabbian, D., Asplund, M., Barklem, P. S., Carlsson, M., \& Kiselman, D.
2009,  A\&A, 500, 1221

\bibitem[2012]{fortney12}
Fortney, J. J. 2012, ApJ, 747, L27

\bibitem[2004]{garnett04}
Garnett, D. R., Edmunds, M. G., Henry, R. B. C., Pagel, B. E. J., \& Skillman, E. D.
2004, AJ, 128, 2772

\bibitem[1991]{hibbert91}
Hibbert, A., Bi{\'e}mont, E., Godefroid, M., \& Vaeck, N. 1991, J. Phys. B, 24, 3943

\bibitem[1993]{hibbert93}
Hibbert, A., Bi{\'e}mont, E., Godefroid, M., \& Vaeck, N. 1993, A\&AS, 99, 179      

\bibitem[2009]{holmberg09}
Holmberg, J., Nordstr\"{o}m, B., \& Andersen, J. 2009, A\&A, 501, 941

\bibitem[2008]{gustafsson08}
Gustafsson, B., Edvardsson, B., Eriksson, K., et al. 2008, A\&A, 486, 951

\bibitem[1999]{gustafsson99}
Gustafsson, B., Karlsson, T., Olsson, E., Edvardsson, B., \& Ryde, N. 1999, A\&A, 342, 426

\bibitem[2003]{johansson03}
Johansson,  S., Litz{\'e}n, U., Lundberg, H., \& Zhang, Z. 2003, ApJ, 584, L107

\bibitem[1999]{kaufer99}
Kaufer, A., Stahl, O., Tubbesing, K., et al. 1999, The Messenger, 95, 8

\bibitem[2001]{kiselman01}
Kiselman, D. 2001, New Astron. Rev., 45, 559

\bibitem[2005]{kuchner05}
Kuchner, M.J., \& Seager, S. 2005, arXiv:astro-ph/0504214

\bibitem[2012]{lind12}
Lind, K., Bergemann, M., \& Asplund, M. 2012, MNRAS, 427, 50

\bibitem[2012]{madhusudhan12}
Madhusudhan, N., Lee, K. K. M., \& Mousis, O. 2012, ApJ, 759, L40
 
\bibitem[2011]{mashonkina11}
Mashonkina, L., Gehren, T., Shi, J.-R, Korn, A. J., \& Grupp, F.
2011, A\&A, 528, A87

\bibitem[2003]{mayor03}
Mayor, M., Pepe, F., Queloz, D., et al. 2003, The Messenger , 114, 20


\bibitem[2012]{nissen12}
Nissen, P. E., \& Schuster, W. J. 2012, A\&A, 543, A28

\bibitem[1983]{olsen83}
Olsen, E. H. 1983, A\&AS, 54, 55

\bibitem[2009]{pereira09}
Pereira, T. M. D., Asplund, M., \& Kiselman, D. 2009,  A\&A, 508, 1403

\bibitem[2011]{petigura11}
Petigura, E. A., \& Marcy, G. W. 2011, ApJ, 735, 41

\bibitem[2009]{ramirez09}
Ram\'{\i}rez, I., Mel\'{e}ndez, J., \& Asplund, M. 2009, A\&A, 508, L17

\bibitem[2006]{skrutskie06}
Skrutskie, M.~F., Cutri, R.~M., Stiening, R., et al. 2006, AJ, 131, 1163

\bibitem[2008]{sousa08}
Sousa, S.G., Santos, N. C., Mayor, M., et al. 2008, A\&A, 487, 373

\bibitem[2005]{takeda05}
Takeda, Y., \& Honda, S. 2005, PASJ, 57, 65

\bibitem[1955]{unsold55}
Uns{\"o}ld, A. 1955, Physik der Sternatmosph{\"a}ren, 2nd ed. (Berlin: Springer Verlag)

\bibitem[2007]{leeuwen07}
van Leeuwen, F. 2007, Hipparcos, the New Reduction of the Raw Data,
(Astrophys. Space Sci. Library, vol. 350; Dordrecht, Springer)

\bibitem[2011]{winn11}
Winn, J. N., Matthews, J. M., Dawson, R. I., et al. 2011, ApJ, 737, L18

\bibitem[2003]{yi03}
Yi, S. K., Kim, Y. -C., \& Demarque, P. 2003, ApJS, 144, 259

\end{thebibliography}
\end{document}